\documentstyle[epsf,prl,aps,twocolumn]{revtex}
\begin{document}
\draft
\title{Gravitational redshifts in electromagnetic bursts occuring
near Schwarzschild horizon}
\author{ Janusz Karkowski $^1$ and}
\author{ Edward Malec $^{1,2}$   }
\address{  $^1$ Institute of Physics, Jagiellonian University, 30-064 Krak\'ow,
Reymonta 4, Poland   }
\address{$^2$ Max Planck Institute for Gravitational Physics AEI, Golm, Germany   }

\maketitle

\begin{abstract}
It was suggested earlier  that the gravitational redshift
formula can be invalid  when the effect of the backscattering is strong.
It is   demonstrated  here numerically,  for an exemplary electromagnetic pulse that
is: i) initially located very close to the horizon of a Schwarzschild black hole
and ii)   strongly backscattered, that  a mean frequency does not obey
the standard redshift formula.  Redshifts    appear to depend on
the frequency and there manifests  a backscatter-induced blueshift in the
outgoing radiation.
\end{abstract}

\date{ }

\section{ Introduction}

Standard  derivations of the gravitational redshift base on
the approximation of geometric optics (see, for a   discussion,
\cite{Sachs} -  \cite{Strauman} ). This problem
has been recently reconsidered  in the wave formulation
\cite{ME2002}. It was  shown, for a class of
compact shocks that  are separated in some sense \cite{sense}
 from the horizon
of a Schwarzschild black hole, that the energy flux
scales accordingly to the redshift formula. Introducing
 the concept of photons and assuming that their number is conserved,
one again arrives at the standard relation  for   the frequency.
As an alternative approach  one can apply  the Fourier
analysis,  as outlined below in Sec. II.

The crucial feature behind the above mentioned compactness and
separateness conditions is that when they hold true, then the backscatter
is negligible \cite{Karkowski}. It was remarked  in \cite{ME2002} that
a compact pulse of radiation  that is exposed to a significant
backscattering must not obey the familiar redshift relation. This
is suggested by the following reasoning. A spatially compact wave
packet consists of a superposition of monochromatic waves of different
frequencies. The effect of the backscattering is stronger  for low
frequency waves than for high frequency ones, that is   spectral
amplitudes of the former are stronger damped than those of the latter.
Therefore one  expects  that the backscattering induces a shift in a
mean frequency, so that it
does not satisfy the gravitational redshift formula. The aim of this
investigation is to extend results of \cite{ME2002} and  find
a numerical example in favour of this conjecture.  This examplary
wave pulse {\bf must break } -- and in fact    {\bf does so},
in accordance with \cite{ME2002}  -- the compactness condition of \cite{sense}.

In the next section we  write   equations and  review relevant
results of the preceding papers \cite{ME2002}, \cite{malec2000}.
Sec. III describes an electromagnetic wave that does not
comply to the standard redshift formula. Sec. IV presents a short summary.

\section{Gravitational redshift: classical  derivation}

The space-time geometry  is defined  by
the Schwarzschild  line element,
\begin{equation} ds^2 =- (1-{2m\over R})dt^2 +
{1\over 1-{2m\over R}} dR^2 +
R^2 d\Omega^2~,
\label{1}
\end{equation}
where $t$ is a time coordinate, $R$ is an  areal radius
and $d\Omega^2=d\theta^2 +\sin^2 \theta d\phi^2 $
(with $0\le \theta \le \pi $ and $0\le \phi <2\pi $)
is the line element on the unit sphere.
The Newtonian gravitational constant $G$
and  the velocity of light $c$ are put equal to 1.
We define the tortoise coordinate $r^*=R+2m\ln ({R\over 2m}-1)$
and  $\eta_R \equiv 1-{2m\over R}$.
We will consider only the dipole electromagnetic term  and,
more specifically, choose  the potential one-form $A=\sqrt{3/2}
\sin^2\theta \Psi (r^*,t)d\phi $.
The unknown function $\Psi $   satisfies the equation  \cite{Wheeler}
\begin{equation}
(-\partial_t^2 + \partial_{r^*}^2)\Psi = \eta_R{2\over R^2}
\Psi .
\label{2}
\end{equation}
Let the electromagnetic strength field tensor be $F_{\mu \nu }$.
The stress-energy tensor of the electromagnetic field
reads   $T_{\mu }^{\nu }=F_{\mu \gamma }
F^{\nu \gamma }-(1/4)g_{\mu }^{\nu }F_{\gamma \delta }
F^{\gamma \delta }$ and the time-like translational
Killing vector is denoted as $\zeta $.  One can
define   the energy flux $\hat P_R$,
\begin{eqnarray}
 \hat P_R(R,t)\equiv {1\over \sqrt{\eta_r}}\int_{S(R)}dS(R)n_rT^{r\mu }\zeta_{\mu }.
\label{2.1}
\end{eqnarray}
Here $n$ is the unit normal to the sphere $R$ and $dS$ is the standard
area element on $S$.
$\hat P_R$ is equal to
\begin{eqnarray}
 \hat P_R(R,t)  =
 -{4\pi \over   \sqrt{\eta_R}} \partial_t\Psi \partial_{r^*}\Psi .
\label{2.5}
\end{eqnarray}
Let  $\tilde \Gamma_{R_0}$ be a null geodesic directed outward
from the point $R_0$ of the initial hypersurface and let
$\tilde \Gamma_{ R_0,  (R,t) }$  be a segment of $\tilde \Gamma_{R_0}$
that connects $R_0$ and $(R,t)$. Comparing the   energy fluxes
through the spheres $S(R)$ (where $R>>2m$) and the initial   $S(R_0)$,
one obtains    \cite{ME2002}
\begin{eqnarray}
  \hat P_R(R) \approx
    \sqrt{\eta_{R_0}\over \eta_R}    \hat P_R(R_0) ,
\label{2.2}
\end{eqnarray}
assuming that the initial support of a dipole wave packet is contained
in the annulus $(a,b)$ such that   $(b-a)/(a\eta_a^5 )<<1$.
Here $a$ and $b$ are the {\bf areal radii}.

There are two ways to derive  the standard   redshift
formula from (\ref{2.2}).

i) {\it Eclectic approach}. The  condition
\begin{equation}
(b-a)/(a\eta_a^5 )<<1
\label{cond}
\end{equation}
implies the validity of the geometric optics approximation.
Thence  one can write $\hat P_R(R_0)= \hbar \hat \omega_{R_0} N_{R_0}$
and $\hat P_R(R)=\hbar \hat \omega_RN_{R}$, where $\hat \omega_{R_0}$
and $\hat \omega_R$ are the mean frequencies of the initial and
final pulses (as measured by static observers)
 and $N_{R_0}$ and $N_R$ are  the respective photon
fluxes. If the photon fluxes are  conserved, then from (\ref{2.2})
one  infers
\begin{equation}
\hat \omega_R=\sqrt{\eta_{R_0}\over \eta_R}\hat
\omega_{R_0}.
\label{2.4}
\end{equation}
ii) {\it Classical approach.}  One can Fourier-analyze  the quantity
representing the strength field tensor, $h=(-\partial_t+\partial_{r^*})\Psi $;
this is preserved along $\tilde \Gamma_{R_0}$, if  condition (\ref{cond}) is satisfied.
Let the spectral strength field  density be $g(\omega_{\tau } )=
\int dt e^{-i\omega_{ \tau }t} (-\partial_t+\partial_{r^*})\Psi  $. Let the
support of $g(\omega )$ be $\tilde \Omega $. If one assumes the normalization
condition $\int_{\tilde \Omega } d\omega_{\tau } |g(\omega_{\tau })|=1$,
then  the average frequency can be defined  as $\hat \omega_{\tau } =
\int_{\tilde \Omega } d\omega_{\tau } \omega_{\tau } |g(\omega_{\tau })|$.
When the analysis is done with respect the asymptotic  time $t$,
then  asymptotic  mean frequency $\hat \omega $ is found.
The Fourier analysis with respect
the proper time $d\tau = \sqrt{\eta_R}dt$ of a static observer located
at $R$  gives a corresponding   mean frequency $\hat \omega /
\sqrt{\eta_R} $. Thus
the two   frequencies satisfy (\ref{2.4}).

\section{Counterexample to the standard redshift formula}

It was pointed out in \cite{ME2002} that (\ref{2.4})
is valid for very  compact initial pulses - those satisfying
the assumption (\ref{cond}).    This   condition
demands not only that a pulse is compact, but also that its relative width
is small  in comparison to the relative distance from the black hole horizon.
It implies that   the
approximation of the geometric optics is valid and that the
backscatter is absent. We conjecture, basing on a numerical evidence,  that this
  can be relaxed to  $ (r^*(b)-r^*(a))/(2m)<<1$.

In the case of  a radiative pulse that is exposed to a significant
  backscattering the relation (\ref{2.4}) would not hold.
Let us again review arguments in favour of this conclusion. A  wave
pulse is superposed from  monochromatic waves
of different frequencies. When resolved spectrally by an observer,
the pulse can be seen as a collection   of peaks, each characterized by
some mean frequency. The effect of the backscattering is
more pronounced for low frequency waves than for high frequency ones
- the spectral  amplitudes of the former are stronger damped than of
the latter. Therefore one would expect that mean frequencies
$\hat \omega_{R_0}, \hat \omega_{R}$ of the
initial and final pulses will not conform to (\ref{2.4}). The observed
mean frequency  $\hat \omega_{R}$ of an outgoing pulse of radiation
can be  blue-shifted  in comparison to
the value  $\sqrt{\eta_{R_0}\over \eta_{ R}}\hat \omega_{R_0}$.
In these circumstances an attempt to fit observed data
of a resolved multi-peak  spectrum to the simple scaling law of
(\ref{2.4}) can lead to redshifts depending on the frequency.

Our aim  is to find numerical solutions corresponding to  compact
initial data, satisfying the  compactness assumption $b-a<<a$, such that:

i)  the spectral amplitude as a function of $\omega $
has  several peaks with well resolved  mean frequencies;

ii) the  evolution exhibits significant backscatter
(hence the energy flux is diminished);

iii)   some of  mean frequencies at $R_0$ and $R$ do not comply
to (\ref{2.4}).

The condition $b-a<<a$ (we remind the reader that $a$ and $b$ are areal radii)
is imposed in order to guarantee that initially
the wavelengths are much smaller than the   radius of the wave front
and the curvature radius, i.e.,
that   the assumption of the geometric optics is satisfied at the emission region.
On the other hand we {\bf do not assume}   the validity
  of (\ref{cond}); our initial data are such that $r^*(b)-r^*(a) = 4\pi m$,
which breaks down this inequality. Let us remark
that with such data one can have the geometric optics approximation
being valid both at the  emission and detection regions, but being evidently
broken around $R\approx 3m$, when the wavelengths and the curvature
radius are of the same order.

Below we present data describing one  of many analysed   examples.
The mass is normalized to unity and the Schwarzschild radius
$R_g$  is equal to 2. Let $I\equiv (r^*(a), r^*(b))$,
the  support of initial radiation, be comprised between
$a=2+1.97\times 10^{-18}$ and $b= 2+10^{-15}$ or, in terms of the tortoise coordinate,
between $r^*(b)\approx -68.46$ and $r^*(a )\approx -68.46- 4\pi $.
The purely outgoing initial data have the form $\tilde \Psi = \partial_tf(r^*-t)
+f(r^*-t)/R$ \cite{malec2000}, where $f$ is a free datum in $I$ but vanishes
 to the left from $a$.
With $I$ being so compact in terms of the areal radius it is reasonable
to ignore  the $R$ - dependence in this expression and to prescribe
initial $\Psi $ as being dependent only on $r^*-t$.
Thus the initial data can be  defined  by   prescribing $\tilde \Psi $
instead of $f$. We choose $\Psi (r^*)=\sin^2 ((r^*-r^*(a)))$ and
$\partial_t\Psi (r^*)=-\sin (2(r^*-r^*(a)))$   for
$r^*\in I\equiv (r^*(a), r^*(b))$ and  $\Psi (r^*)=\partial_t\Psi
=0$ outside $I$.

Notice that the size of the support of the initial pulse, $i=4\pi $,  satisfies
the inequality $i>2$, which is the plausibility condition (according to
the foregoing conjecture)  for  a significant backscatter.

   The signal  is measured during the time
interval $32 \pi $ at two observation
points, $r^*_{0}=-68.44$  and $r^*_1= 199.16$; in terms of the
areal radius we have $R_0=2+1.012\times 10^{-15}$ and $R=190.07$.
The grid was $120000\times 60000$.
Two snapshots of the evolving configuration taken at the
above  pair of observation
points  are presented  in Fig. 1.
\begin{figure}[1]
\epsfxsize=6cm
\centerline{\epsffile{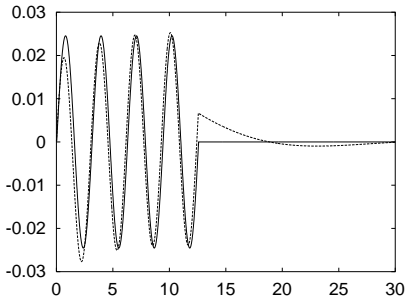}}
\caption{ Temporal dependence of $h\equiv (-\partial_t+\partial_{r^*})\Psi $
(with values of $h$  put on the ordinate)
as observed at $R_0=(2+10^{-15})m$ (solid line) and at $R=190.07m$ (broken line).
The time (with values put on the abscissa) is measured from the moment
 of the arrival of the wave front at each of the
observation points. The   part of the radiation   seen at $R$ (broken line)
after  $t=4\pi m$ consists solely of the backscattered radiation.
The signal is negligible after $t>30m$ - its amplitude is much smaller than that
of the oscillatory part.}
\end{figure}

  The main pulse passes through $R$ in less than
$4\pi $ (in units of $m$) while the scattered wave appears later and
quickly vanishes, becoming negligible after $t=30 m$, when the tail dominates.
The decay of the tail asymptotics  should be like $t^{-5}$,
according to  \cite{Price}. This asymptotic exponent was in fact obtained
in many trial runs of our  numerical code with  initial pulses
located outside $a=3m$ \cite{expla}.

 Figure 2
in turn shows the spectral decomposition of $h=(-\partial_t+
\partial_{r^*})\Psi    $, obtained by employing  the FOURCO package
(netlib).  Fast Fourier Transform  (IMSL library) was also used,
with the purpose of checking the numerical results coming from the
application of FOURCO. We depicture
{\bf normalized frequencies} in order to see clearly the frequency shift
that is caused by the backscatter. The spectrum that is seen at $R_0 $
(solid line) is rescaled by $\sqrt{\eta_{R_0}}$  and the
spectrum seen at $R$ (broken line) is rescaled by the factor
$\sqrt{\eta_R}$.   In the case of negligible
backscattering both rescaled spectra would have to coincide.

Figure 2 shows the spectral amplitudes $|g(\omega )|$ in function of the
frequency $\omega $. There  is a number of peaks, seen at
both observation points. We calculated average frequency for
each of the resolved peaks, by choosing as the support $\tilde \Omega $
of each pulse the interval between the consecutive minima of
$|g(\omega )|$. In the case of the lowest frequency peak one can observe
a 25-percent blueshift, from 0.25 to 0.31, with variances 0.1
and 0.08, respectively; the broken and solid lines of Fig. 2
coincide for $\omega >0.5/m$. In the remaining peaks the effect is
seemingly absent. This  can be interpreted as demonstrating that  this
 backscatter-induced blueshift is frequency dependent.
$1/(4m)$ is approximately the asymptotic frequency of the quasinormal modes.
It can be seen from Fig. 2 that the effect of the
 backscattering strongly weakens
  modes having frequencies $\omega_b$ (as observed by an observer
static at $b$), such that $\omega_a\sqrt{\eta_b}$ (which
would be the frequency detected by an  asymptotic static observer)
 is smaller than  $1/(8m)$ - a half of  the asymptotic frequency of quasinormal modes.
\begin{figure}[2]
\epsfxsize=6cm
\centerline{\epsffile{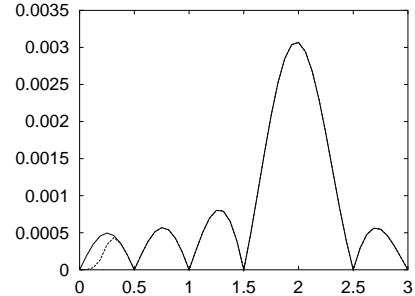}}
\caption{Frequency amplitude $|g(\omega )|$ (depictured on the ordinate)
as observed at $R_0=(2+10^{-15})m$ (solid line) and at $R=190.07m$ (broken line).
The {\bf normalized frequencies} (see the text above)
 are put on the x-axis  (scaled in units of $1/m$). }
\end{figure}

Modes with $ \omega_b\sqrt{\eta_b} >1/(2m)$ are    unchanged, while those
in the interval $(1/(8m)<\omega_b\sqrt{\eta_b}< 1/(2m)$
 undergo a damping of the spectral amplitude,
which is  the stronger the smaller  the frequency.   The physical reason
for this is following.  Our initial wave pulse has to penetrate through the
potential barrier having maximum at $r=3m$; the  modes with frequencies
close to those of quasinormal modes are much stronger scattered and their
transmission coefficient is small.

Let us point out that the backscatter is negligible -- due to the
weakness of the effective potential  --
as long as a wave pulse remains in a region
characterized by the tortoise variable $r^*<<-1$.
Therefore    the standard redshift formula applies in the
 zone $r^*<<1$.
Define a shifted  radiation pulse as follows:   $\Psi (r^*+D)=\Psi(r^*)$
for  $D<0$). Notice that   this translated pulse   will possess
a {\bf different}  frequency profile. Let $b_D$ be defined by the equation $r^*(b)+D=
b_D+2\ln (b_D/(2m)-1)$ and let $(w_b)$ be a frequency measured by a
static observer located at $r^*(b)$.
A static observer located at $r^*(b)+D$ will   measure a blueshifted frequency
 $(w_b)\sqrt{\eta_b/\eta_{b_D}}$.
     On the other hand {\bf a static observer located at infinity
would detect the same set  of frequencies irrespective of $D$}.
   Moreover, this observer would notice that the
 deviation  from the redshift formula (if that can be seen  in the detected
frequency range) does not depend on the position of the initial pulse.

If however a {\bf physical} source   of a wave
is fixed  and shifted
 from $r^*(b) <<-1$ by some $D<0$  then
 the two local sets of frequencies $(\omega_{bi}, \omega_{b_Di}$)
measured by two (shifted  by a distance $D $) local static observers {\bf do coincide},
$\omega_{bi}=\omega_{b_Di}$. (We assume that tidal effects can be ignored.)
The    static observer at infinity would then see:
i)  two {\bf different sets of redshifted frequencies},  $(\omega_{bi,\infty })$
and $(\omega_{b_Di,\infty })$, with   $\omega_{bi,\infty }=\sqrt{\eta_b}\omega_{bi}$
and $\omega_{b_Di,\infty }=\sqrt{\eta_{b_D}}\omega_{b_Di}$
for all frequencies  which satisfy conditions
  $\omega_{bi,\infty }>>1/(4m) $ and $\omega_{b_Di,\infty }>>1/(4m)$;
   ii) those frequencies,   for which
 $\omega_{bi,\infty }\le 1/(4m)$ and/or  $\omega_{b_Di,\infty }\le 1/(4m)$, would
 disobey the   standard redshift formula.  It is clear  that for any fixed wave
 source with frequencies $\omega_i$ one can find a location  $b$ close to the
 black hole horizon such that some of
(the would-be asymptotically detected) frequencies $\eta_b\omega_i$ are of the  order
of  $1/(4m)$, so that (according to our numerical example) one could see a failure
of the standard redshift formula.

If that numerically discovered  phenomen is generic, then there  would
exist a natural cutoff (of the order of  $1/(4m)$) for the asymptotically
observed frequency of those wave pulses that are close to an event horizon.
A wave  of a  mean frequency $\omega_b$
satisfying the condition  $\omega_b\sqrt{\eta_a}<< 1/(4m)$
would not be observed   by a distant observer.
In the light of the above   the heuristic
 analysis that was reported in the
beginning of this section, should be probably supplemented by adding
the following: {\it  if a wave peak possesses a contribution
with frequencies much smaller than   $1/(4m)$ then (and only then) its
mean frequency will be influenced by the backscatter}.

\section{Final remarks}

The backscattering can modify the relation between initial and asymptotic
frequencies. A characteristic  frequency  (circa $1/(4m)$ in the hitherto
used units) that appears useful in this
context is (in standard units) $f_c\approx 0.25\times 31484\times M_0/m$ Hz,
where $M_0$   is the solar mass. Let us remark that $f_c$ is the
frequency  typical for  the quasinormal modes of the electromagnetic
radiation propagating in the Schwarzschild spacetime.

A numerical example of Sec. III reveals a robust difference between
the standard redshift prediction    (\ref{2.4}) and the actual observation.
It is observed that for a radiation pulse having an asymptotic frequency
$f_{\infty }\approx f_c$ the initial frequency $f_R$ is
noticeably smaller than $f_c/\sqrt{\eta_R}$, a number following
from the redshift formula (\ref{2.4})). The  peaks with frequencies $f_{\infty }>>f_c$
are seemingly not influenced by the backscattering. This probably
is not always true for all radiation pulses having  mean frequencies
much bigger than $f_c$,
but we believe that in the latter case the deviation from the redshift
formula must be small.  Rephrasing this fact in terms of initial
data, it is  probably fair to say that $(r^*(b)-r^*(a)<<2m$ is
the sufficient condition  for the validity of the redshift formula (\ref{2.4}).
That is a much less stringent condition than that used in
\cite{{ME2002}}.
Let us stress that in the case of astrophysical black holes of stellar
origin   the deviation from the  law (\ref{2.4}) can be seen
  by an asymptotic observer in the  radio part  of the electromagnetic
spectrum, with wavelengths of the order of  100 km; we do not claim that
this phenomenon is of astrophysical interest.

Let us point out that if  the  backscattering is negligible then:
 i) all of the energy of an outgoing pulse
would get  to infinity; ii) its energy flux would be diminished,
according to  (\ref{2.5}). The only mechanism  that can imply
the loss of the radiation energy is through the backscatter of the
radiation  off the curvature of the spacetime. This is
clearly shown in the wave formulation \cite{malec2000}, but a consistent
quantum field  theoretic treatment (in the curved Schwarzschild spacetime)
 should give the same answer.

 In the present paper the consideration is focused  only on the dipole
term, but a similar analysis with the same conclusions can
to be done in any multipole order.
An analogous phenomenon, leading to the demodulation of  gravitational
signals, will also manifest in the spectra of gravitational waves;
this in principle may be detected.

Acknowledgments.    This work has been suported
in part  by the KBN grant 2 PO3B  006 23.


\begin{references}
\bibitem{sense} It is meant here   that $(b-a)/(a(1-2m/a)^5 )<<1$,
where the areal radii $a$ and $b$ are the inner and outer boundaries
 of the initial wave pulse.
\bibitem{expla} The integration time, that would be needed in order to obtain the
asymptotic tail behaviour in the case of the initial data  of Sec.  III,
exceeds by far the integration time required in order to solve our task.
\end{references}
\end{document}